\documentclass[12pt,preprint]{aastex} 

\usepackage{textcomp}
\usepackage{graphicx}
\usepackage{natbib}
\usepackage{float}
\usepackage{bookmark}
\bibpunct{(}{)}{;}{a}{}{,}

\usepackage{txfonts}
\usepackage{color}
\newcommand\ionn[2]{#1$\;${\scshape{#2}}}%
\newcommand{\varcsec}{^{\prime\prime}}
\renewcommand{\arcsec}{.\hspace{-0.9mm}'\!\hskip0.4pt'\hspace{-0.2mm}}

\slugcomment{Received ........; Accepted .........}

\shorttitle{Convection in Penumbra}

\shortauthors{Jayant Joshi et al.}

\begin{document}

\title{Convective Nature of Sunspot Penumbral Filaments: Discovery of Downflows in the Deep Photosphere}

\author{Jayant Joshi$^{1,2}$, A.~Pietarila$^{1,3}$, J. Hirzberger$^{1}$, S. K. Solanki$^{1,4}$, R. Aznar Cuadrado$^{1}$ and L.~Merenda$^{1}$ }
\affil{$^1$ Max-Planck-Institute f\"ur sonnensystemforschung, Max-Planck-Str. 2, 37191 Katlenburg-Lindau, Germany}
\affil{$^2$ Institut f\"ur Geophysik und Extraterrestrische Physik, Technische Universit\"{a}t Braunschweig, Braunschweig, Germany}
\affil{$^3$ National Solar Observatory, 950 N. Cherry Avenue, Tucson, AZ 85719, USA}
\affil{$^4$ School of Space Research, Kyung Hee University, Yongin, Gyeonggi Do, 446-701, Korea}
       \email{joshi@mps.mpg.de}
\begin{abstract}
We study the velocity structure of penumbral filaments in the deep photosphere to obtain direct 
evidence for the convective nature of sunspot penumbrae. A sunspot was observed at high spatial 
resolution with the 1-m Swedish Solar Telescope in the deep photospheric \ionn{C}{i}\,5380\,\AA\ 
absorption line. The Multi-Object Multi-Frame Blind Deconvolution (MOMFBD) method is used for image
restoration and straylight is filtered out. We report here the discovery of clear redshifts in the
\ionn{C}{i}\,5380\,\AA\ line at multiple locations in sunspot penumbral filaments. For example, 
bright head of filaments show larger concentrated blueshift and are surrounded by darker, redshifted
regions, suggestive of overturning convection. Elongated downflow lanes are also located beside 
bright penumbral fibrils. Our results provide the strongest evidence yet for the presence of 
overturning convection in penumbral filaments and highlight the need to observe the deepest 
layers of the penumbra in order to uncover the energy transport processes taking place there.
\end{abstract}

\keywords{Sun:photosphere -- Sunspots -- Convection}

\section{Introduction}

In recent years, indirect evidence for the presence of convection in sunspot penumbral filaments 
has been growing (e.g., Scharmer 2009). E.g., the twisting motion of penumbral filaments 
is taken as a signature of overturning convection (Ichimoto et al. 2007; Zakharov et al. 2008; Spruit
et al. 2010; Bharti et al. 2010). Using high resolution spectropolarimetric observations, Zakharov
et al. (2008) estimated that such
motions can provide sufficient heat to the surface layers of the penumbra to explain its relatively
high brightness. The correlation of the apparent velocity of the twisting motion with the local 
brightness of the filaments obtained by Bharti et al. (2010), supports convection as a major source 
of heat transport in sunspot penumbral filaments. Overturning convection in penumbral filaments 
is a natural and important feature in three-dimensional MHD simulations of sunspots (Rempel et al. 2009a,b; Rempel 2011)
. In particular, Rempel et al. (2009b) found upflows along the central axes and 
downflows at the edges of the filaments. Direct observational evidence for this scenario is, however, 
so far missing, because downflows have not been measured in the body of a penumbra, although the 
twisting filaments provide indirect support. The simulations indicate that the convective structures
and motions are restricted to the subsurface and surface layers. Since most spectral lines used 
for diagnostic purposes sample mid-photospheric heights, this may explain why it has not been 
possible to obtain direct evidence of overturning convection in penumbral filaments 
(see, e.g., Franz \& Schlichenmaier 2009; Bellot Rubio et al. 2010). In particular, downflows at the sides of 
penumbral filaments have not been reported.   

In this study we probe the deep layers of the photosphere in search of such downflows by analyzing
high resolution observations in \ionn{C}{i}\,5380.3\,\AA, obtained at the 1-m Swedish Solar Telescope. 
In the quiet Sun, \ionn{C}{i}\,5380.3\,\AA\, has a mean formation height of around 40 km above the
continuum optical depth $\tau_c = 1$\ at 500 nm (Stuerenburg \& Holweger 1990) making it ideal for this purpose.

%
%

\section{Observation and Data Reduction}
 
We observed a decaying sunspot with a one-sided penumbra in active region NOAA~11019 
(cf. Fig. 1) with the 1-meter Swedish Solar Telescope (SST) on 2009, June 02. 
The center of the field of view (FOV) was located at $\mu = 0.84$ (heliocentric angle =
32.7\textdegree). During the observations the seeing conditions were good to excellent, 
with only few interruptions by poorer seeing.

We carried out consecutive spectral scans of the photospheric \ionn{C}{i}\,5380.3\,\AA ,
\ionn{Fe}{i}\,5250.2\,\AA\ and the chromospheric \ionn{Ca}{ii}\,8542\,\AA\ spectral 
lines using the CRISP imaging spectropolarimeter. Here we analyze only the 
\ionn{C}{i}\,5380.3\,\AA\ line\footnote{The \ionn{Fe}{i}\,5250.2\,\AA\ line is used only to
identify "quiet Sun" locations, where the polarization signal in this line is below a given 
threshold.}. Two liquid crystal variable retarders, used to modulate
the light beam, and two 1k$\times$1k-pixel Sarnoff CCD cameras, mounted behind a 
polarizing beam splitter, were used to record the full Stokes vectors at each spectral 
position. A third CCD camera was used to record broad-band images. All three cameras were
synchronized in order to allow post-facto image restoration. We recorded the 
\ionn{C}{i}\,5380\,\AA\ line at 8 wavelength positions ($\lambda - \lambda_0 = 
[-300,-120,-80,-40,0,40,80,120]$\,m\AA). Scanning the line in all four Stokes parameters
required 14 \,s for the \ionn{C}{i}\,5380.3\,\AA\ line. The cadence of these observations,
including the \ionn{Fe}{i}\,5250.2\,\AA\, and \ionn{Ca}{ii}\,8542\,\AA\, scans (not 
considered in the present study) is 29\,s.

\begin{figure}[t]
\centering
      \includegraphics[width=0.65\textwidth]{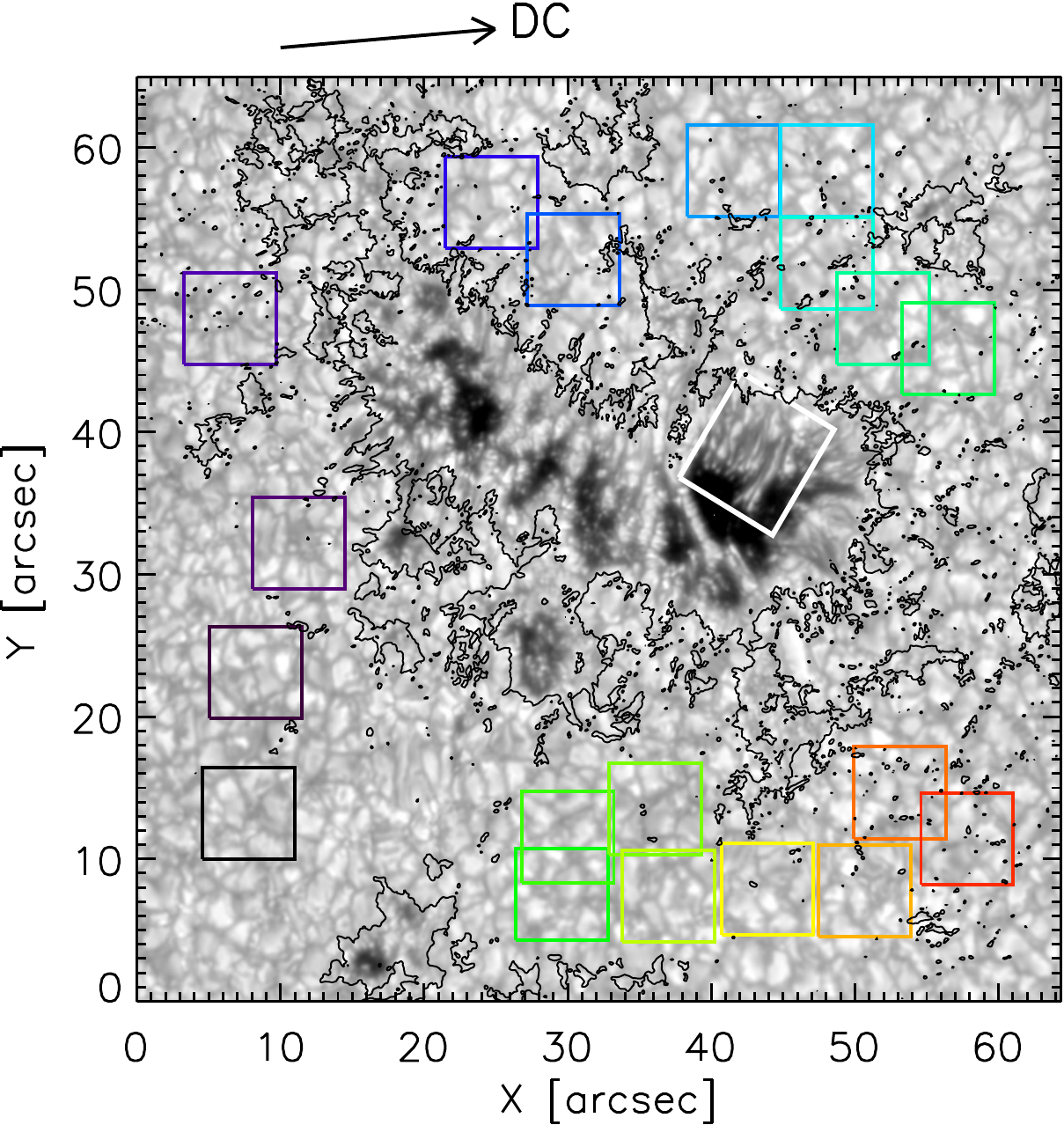}
      \caption{Continuum image at 5380\,\AA\  of the decaying sunspot with a 
               penumbra mainly on one side. Black contours outline regions 
               where the absolute circular polarization averaged over the two
               line wings ($\pm$ 40\,m\AA\, from average line center) of the 
               \ionn{Fe}{i}\,5250\,\AA\ line is greater than 4 \%. Colored boxes 
               ($6\varcsec\times 6\varcsec$) show the different quiet Sun fields 
               used to calculate the velocity reference value. The thick white rectangle 
               marks the portion of the image shown in Fig. 5 in detail.}
\end{figure}

\begin{figure}
\centering
      \includegraphics[width=0.60\textwidth]{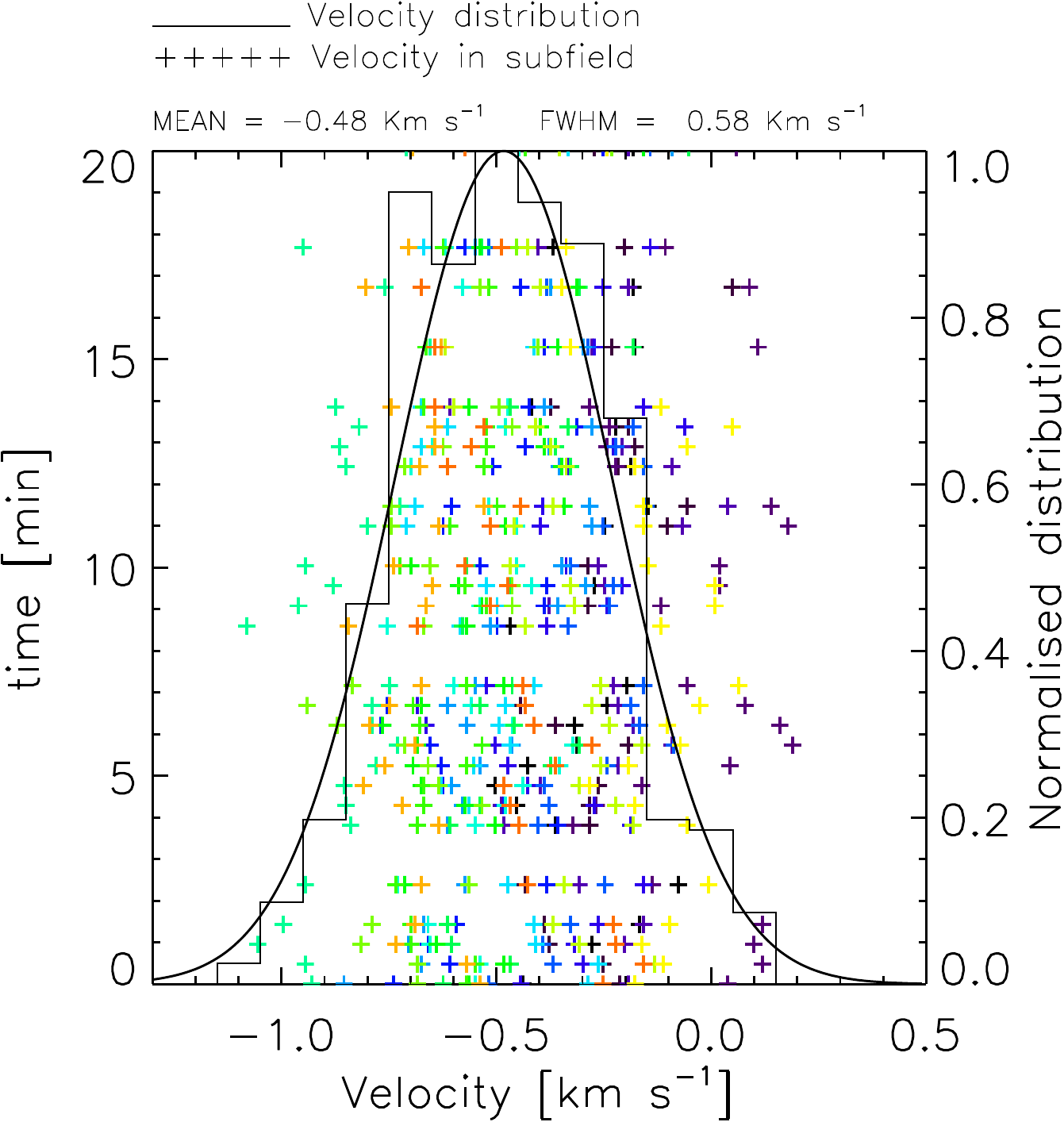}
      \caption{Mean velocities obtained from average line profiles of the subfields
               shown in Fig. 1 (abscissa) vs. time of observation relative to the time
               of the first recording (ordinate). Colors correspond to the subfields bounded by 
               squares of with the same colors in Fig. 1. The velocity distribution and 
               a Gaussian fit are represented by solid lines.}
\end{figure}

\begin{figure}
\centering
      \includegraphics[width=0.60\textwidth]{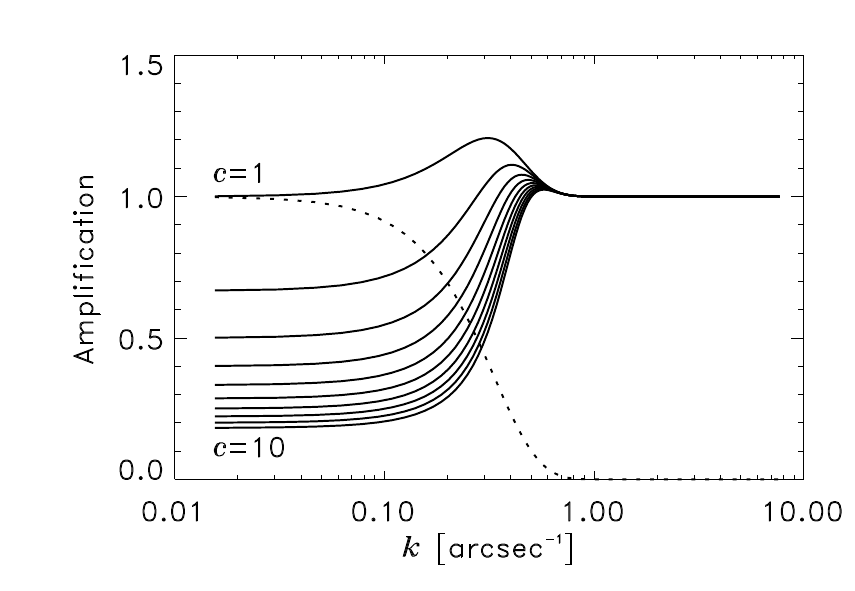}
      \caption{Applied filter to deconvolve straylight. The solid curves denote filter 
               shapes for $c = 1 . . . 10$ (see Eq. (1)). The dotted curve represents 
               the $MTF$, i.e., the Fourier transform of a Gaussian of $1.6\varcsec$ width.
               }
\end{figure}

\begin{figure*}
\centering
      \includegraphics[width=1.0\textwidth]{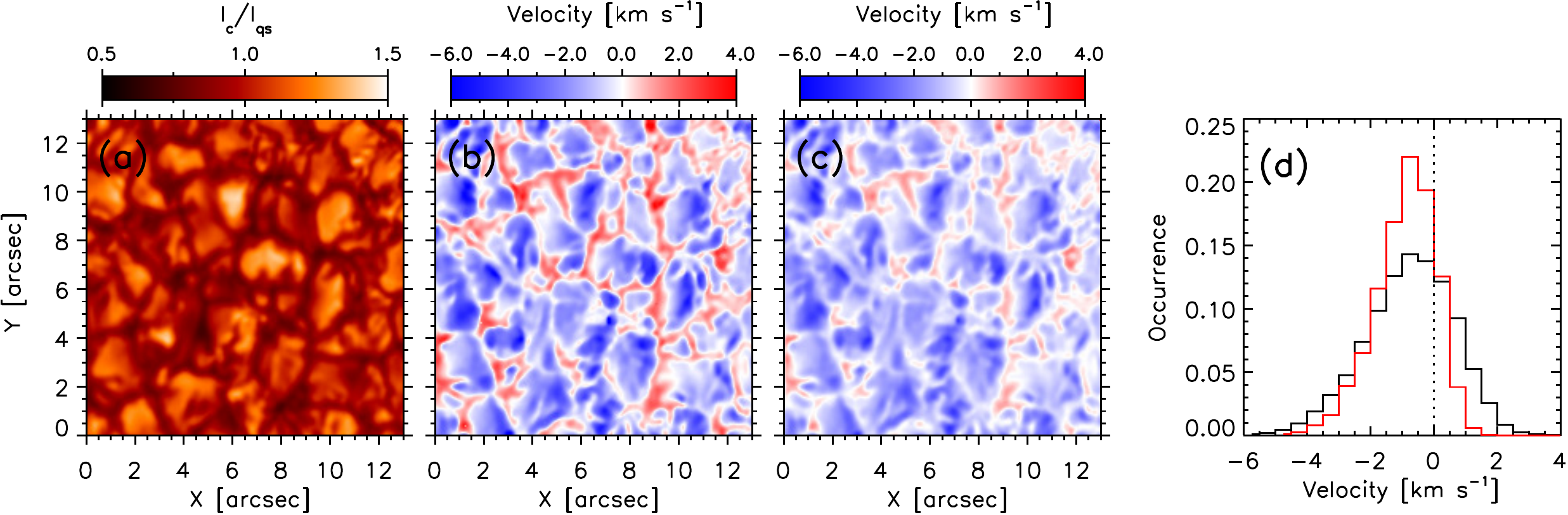}
       \caption{Panels (a) and (b) display a quiet Sun continuum map near \ionn{C}{i}\,5380.0\,\AA\,
                and the corresponding Doppler map after straylight removal, respectively. Panel (c) shows,
                for comparison, the Doppler map before straylight removal while histograms of Doppler 
                velocity before (red) and after (black) straylight removal are plotted in panel (d)}
\end{figure*}

To achieve near diffraction limited spatial resolution ($0\arcsec14$), images were 
reconstructed using the Multi-Object Multi-Frame Blind Deconvolution (MOMFBD) method
(van Noort et al. 2005; L\"{o}fdahl 2002). Images observed in \ionn{C}{i}\,5380\,\AA\ and 
\ionn{Fe}{i}\,5250\,\AA\ were aligned with sub-pixel accuracy by cross-correlating 
the corresponding continuum images. We determined Doppler velocities by fitting a Gaussian
function to the \ionn{C}{i} line. Due to the low Land\'{e} factor ($g_{\rm eff} = 1$,
Solanki \& Brigljevic 1992) and large thermal width of the line, this is an acceptable approach even in a sunspot.
The good quality of the Gaussian fit throughout the penumbra suggests that the line is unblended there, 
unlike in the umbra (see Sect. 3). Another reason for expecting that the line is unblended 
(or at the most rather weakly blended) in the penumbra is that the line strength increases strongly with 
temperature, a nearly unique property of \ionn{C}{i}\, lines among photospheric spectral lines.

The resulting velocity maps show variations of the mean quiet Sun velocities across the 
FOV caused by the absolute wavelength positions varying across the FOV due to 
cavity errors of the etalons of CRISP. These wavelength shifts are smeared out in a continuously 
changing manner due to varying seeing conditions. The applied calibration data (flat fields),
which are well defined in pixel space, are no longer connected to pixel space of the science 
data after restoration (Schnerr et al. 2010). Therefore, the cavity errors cannot be fully 
removed with the flat fields. The absolute wavelength shifts caused by the cavity errors 
are wavelength dependent and the residuals after correction for flat-field cavity shifts are
much higher in the 5380\,\AA\ band than in the 6300\,\AA\ and 6560\,\AA\ bands, used in 
previous studies carried out with CRISP (see, e.g., Scharmer et al. 2008;
Rouppe van der Voort et al. 2010; Ortiz et al. 2010). 
Since the residual wavelength shifts across the FOV are expected to be large-scale, however, they do 
not rule out a study of penumbral fine structure. In order to determine a confidence level of the 
resulting absolute velocity reference point of the Doppler velocities we selected 19 quiet 
Sun subfields of $6\varcsec\times6\varcsec$ size throughout the 
FOV, indicated by the colored squares in Fig. 1. The relative quietness of the 
selected subfields was assured by a threshold value (4 \%) of spectrally averaged absolute
circular polarization\footnote{A full polarimetric model of the SST in the observed spectral
  range is not available so far. Therefore, we are not able to carry out a complete correction
  for instrumental cross-talk and the displayed polarimetric information can be considered 
  only as a rough estimate of the magnetic field structure in the FOV.} in the line wings 
($\pm$ 40\,m\AA\, from average line center) of \ionn{Fe}{i}\,5250.2\,\AA. We calculated mean Doppler
velocities of \ionn{C}{i}\,5380\,\AA\, by spatially averaging the line profiles in each subfield
and repeated this procedure for all the observed sequences, i.e., over 20 min. The results from this 
procedure are shown in Fig.~ 2. The velocity in each subfield fluctuates randomly, but 
does not exhibit a systematic variation with time. In order to get a mean quiet Sun velocity reference, 
we fitted a Gaussian to the velocity distribution and obtained a mean value of $-$480\,m\,s$^{-1}$ 
and a full width at half maximum (FWHM) of 580\,m\,s$^{-1}$ for the \ionn{C}{i}\,5380\,\AA\ line,
 i.e., the residual uncertainty ($1\sigma\,$) in the absolute Doppler velocities is  
$\pm$248\,m\,s$^{-1}$. After applying this procedure, the mean values of the obtained velocity 
maps are calibrated for a convective blueshift of $-$922\,m\,s$^{-1}$ for \ionn{C}{i}\,5380.3\,\AA\
(de La Cruz Rodr\'{\i}guez et al. 2011). For the further analysis we  selected the scan of \ionn{C}{i}\,5380.3\,\AA\, 
made under the best seeing conditions.

\subsection{Straylight Correction}

Ground-based observations are strongly affected by straylight, so that, e.g., 
contrasts are significantly lower than in data recorded above the Earth's atmosphere 
(see, e.g., Danilovic et al. 2008; Hirzberger et al. 2010). We also expect the velocity to be reduced through straylight. The 
amount of straylight and the shape of the corresponding straylight point spread function is 
strongly dependent on the seeing conditions and on the instrument design. The straylight point 
spread functions can be approximated from fitting observations of the solar limb 
(aureolas, see, e.g., Sobotka et al. 1993) or from fitting the profiles of planetary limbs during 
transits (see, e.g., Bonet et al. 1995; Mathew et al. 2009; Wedemeyer-B\"{o}hm \& Rouppe van der Voort 2009). Since we do not have these auxiliary 
data, we applied a rough estimation, assuming that the biggest amount of the straylight stems from 
regions within $1.6\varcsec$ around the respective positions (approximately the tenfold of the spatial
resolution of the data), i.e., we approximated the straylight point spread function with a Gaussian of 
$1.6\varcsec$ width. For deconvolving the images we
used the Wiener filter given in Sobotka et al. (1993) and applied a slight modification, so that it has
the shape
\begin{equation}
 {F(k)=\frac{1}{c} \frac{MTF(k)+1}{MTF(k)^2+{1/c}}} \,,  
\end{equation} where MTF is the modulation transfer function (the modulus of the Fourier transform of the point
spread function), $k$ is the spatial wavenumber and $c$ is a free parameter which defines the straylight
contribution that has to be deconvolved. $F(k)$ for different choices of $c$ is plotted in Fig. 3. 
The advantage of the present form of the filter compared
to the one used in Sobotka et al. (1993) is that it converges to unity for large $k$, i.e., it does not affect
the small-scale structures of the images.

In the present study the value of parameter $c$ has been chosen equal to 2.0, so that the resulting quiet Sun rms image
contrast for the continuum point (at $\lambda = \lambda_0 - 300$\,m\AA\,) of the best scan increases from 
8.5\% in the original image to 13.0\% in the deconvolved image, which is closer to, but still somewhat smaller 
than the contrast obtained from data at a similar wavelength that are much less contaminated by straylight 
(continuum at 5250.4~\AA\, in IMaX/Sunrise observations, see Mart\'{\i}nez Pillet et al. 2011). We finally used this conservative choice, 
although we also tested other straylight functions, including broad Gaussians with
widths up to $16\varcsec$. The results of using stronger straylight removal (larger $c$) or broader Gaussians, 
remained similar, but provided stronger downflows and, for larger $c$ , also stronger contrast. 
Stronger straylight removal also led to bigger scatter in the velocities and somewhat more distorted 
line profiles, which was another reason to keep to the conservative value of straylight.

We selected a $13\varcsec\times13\varcsec$ field in the quiet Sun to compare velocities before and after
straylight removal. Results are shown in Fig. 4. The
redshift in inter-granular lanes increases disproportionately through straylight removal 
(see panels (b), (c) and (d) of Fig. 4. Downflows are particularly affected by straylight because 
they are narrower and are present in darker features. The fact that the downflows remain weaker than 
upflows also after our standard straylight removal confirms that we have been conservative in the removed 
amount of straylight. 

\begin{figure*}
\centering
      \includegraphics[width=0.8\textwidth]{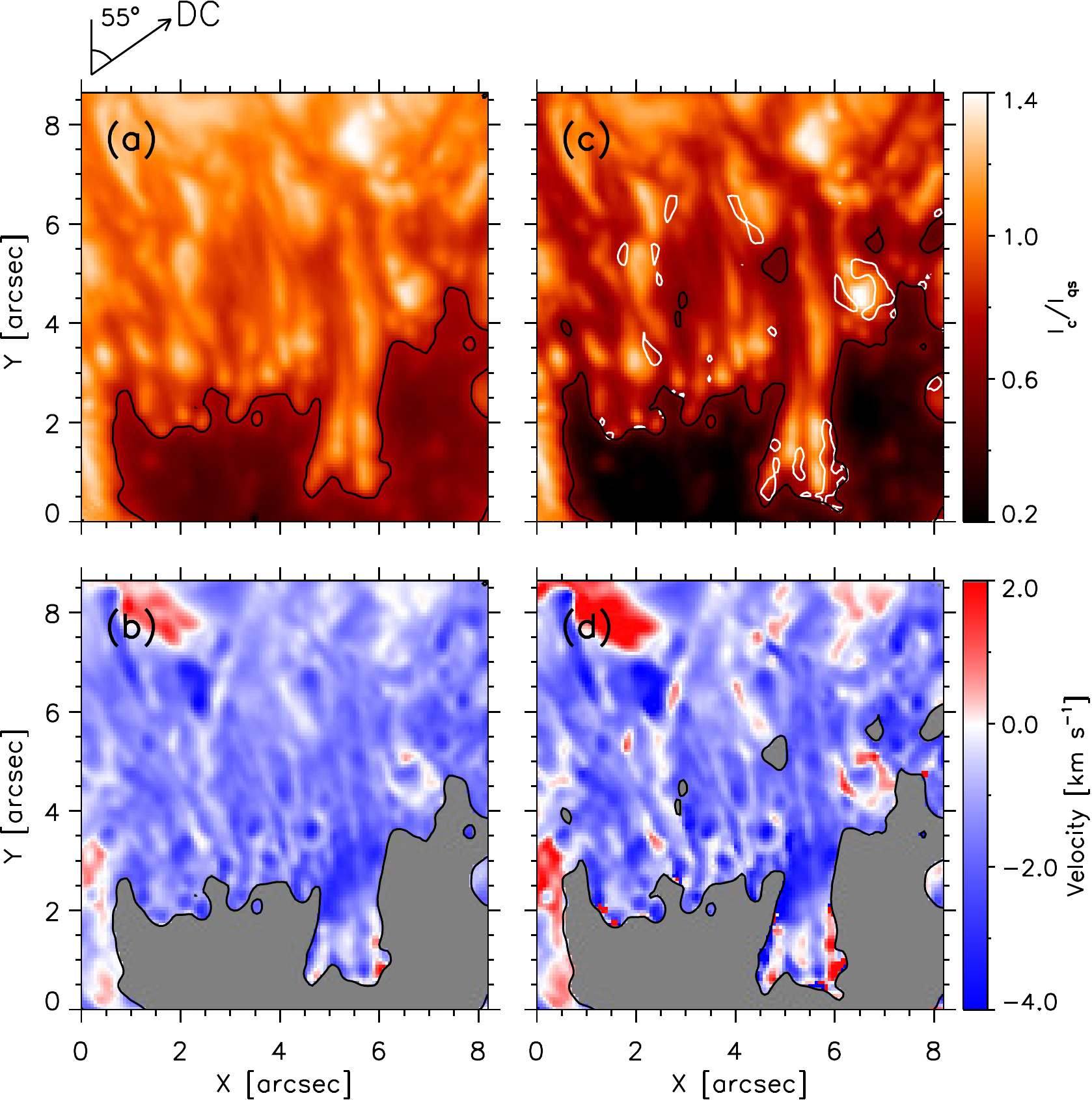}
      \caption{(a) Continuum intensity map in the penumbral region of the observed 
        sunspot obtained from the \ionn{C}{i}\,5380\,\AA\, line. The displayed region 
        is marked by the thick white rectangle in Fig. 1. (b) the  corresponding
        Doppler velocity, (c) continuum intensity map after straylight removal, (d) Doppler 
        velocity after straylight removal. Gray color encircled by black contours 
        in panels (b) and (d) represent areas where continuum intensity is less than 
        $0.6 \,I_{\rm{qs}}$. White contours in panel (c) outline penumbral downflows and show that
       these are located in darker parts of the penumbra. The arrow above panel (a) indicates the 
        direction toward solar disk center.    
        }
\end{figure*}

\section{Results}

Panels (a) and (c) of Fig. 5 show a portion of the observed penumbra as seen in the continuum
intensity (at $\lambda=\lambda_0 - 300$ m\AA), before and after straylight removal, respectively. The 
corresponding maps of Doppler velocity are plotted in panels (b) and (d). Obviously both blue- and 
redshifts are present in the penumbra, in particular after straylight removal. A striking feature in 
Fig. 5b is the localized patches of strong blueshift, up to 3.3 \,km\,s$^{-1}$, coinciding 
with the bright heads of penumbral filaments. These bright heads are surrounded by lanes of gas nearly 
at rest or slightly redshifted. In panel (b) significant redshifts are visible only at two locations, once at the 
side of a filament reaching into the umbra (at $x = 6\varcsec$ and $y = 1\varcsec$) and once, rather 
weakly around the bright and strongly blueshifted head of a filament (at $x = 6.5\varcsec$ and $y = 4.5\varcsec$).

A number of new redshifted patches are found in the Doppler map after straylight 
removal (panel d), appearing dominantly at locations previously seemingly at rest in panel (b). We have grayed out areas
where intensity is below 0.6 times that of the quiet Sun, because $\chi^2$ values of the Gaussian fits to the line
profiles are high in these areas since the line is very weak there and possibly blended (as suggested by the fact that
we obtain mainly strong blueshifts in the umbra contrary to all previous studies based on other spectral lines). 
Redshifts (largest value 2.0 \,km\,s$^{-1}$) show a tendency to be located in dark regions, as can be judged by 
considering the white contours in panel (c). These contours outline the redshift of panel (d). Redshifts are now found
clearly around the head of multiple filaments (e.g., at $x=2\varcsec$ and $y=3\varcsec$ and around the filaments 
protruding into the umbra). Narrow redshifted areas are also found in the middle and outer penumbra beside and
between bright filaments. 

Clearly, in the lower photosphere redshifts are present at many different locations in the penumbra. 
We expect that only a part of redshifted features actually present in the penumbra has been detected.

\section{Discussion}
We have provided the first direct measurements of downflows (reaching 2.0 \,km\,s$^{-1}$) in the body of 
a penumbra. The studied sunspot, located at $\mu =0.84$, had only a partial penumbra, on the disk-center 
side of the spot, so that all well-defined filaments 
partially point to disk center. Hence the Evershed flow contributes a blueshift, so that redshifts in 
Fig. 5d must be caused by downflows (or inflow, which appears rather unlikely, however).  

The \ionn{C}{i}\,5380\,\AA\, line reveals a highly structured velocity pattern, with large variations in
velocity around the head of filaments. Typical blueshift of 3\,km\,s$^{-1}$ found 
in the bright head of the filaments, surrounded by gas displaying redshift (after straylight removal) or 
no shift. We interpret them as strong localized upflows of hot gas in the head of the filaments  following 
Rimmele \& Marino (2006),, but cannot rule out that they are due to Evershed flow. A fraction of this
gas starts moving along the axis of the filaments and forming the Evershed 
flow (Scharmer et al. 2008). The rest of the gas moves to the sides of the filaments and flows downward.
Further downflows are found alongside bright filaments in the middle and outer part of this penumbra. Such
downflows have been predicted by models of penumbral convection (e.g., Rempel et al. 2009a,b; Scharmer \&
Spruit 2006). Earlier observational studies have provided only indirect evidence for such downflows
(Ichimoto et al. 2007; Zakharov et al. 2008; Bharti et al. 2010; Spruit et al. 2010).

The velocity at the heads of the filaments reaches values up to 3.3\,km\,s$^{-1}$, interpreted here as upflows, 
agrees with earlier observational results (Rimmele 1995; Rimmele \& Marino 2006; Hirzberger \& Kneer 2001; Hirzberger et al. 2005). Such upflows
are also consistent with the moving flux tube model presented by Schlichenmaier et al. (1998), which predicts an 
upflow of 4\,km\,s$^{-1}$ at the footpoints of the penumbral filaments, but equally with the interpretation of 
Scharmer et al. (2008) that the Evershed flow is a horizontal flow component of overturning convection. In recent 
years, convection has become an important candidate to explain the heat transport in penumbrae. Based on the 
observed 1\,km\,s$^{-1}$ upflow at the axis of the filament in the upper layers of the photosphere, 
Zakharov et al. (2008) estimated the heat transport by convection in the penumbra to be sufficient for maintaining
the brightness of the penumbra. The upflows up to 3.3\,km\,s$^{-1}$ in the bright heads of penumbral filaments
with downflows at the filament sides, found in the present study, and the coincidence of the strongest upflows with
bright filaments and downflows with dark filaments strongly support that the heat transport in the penumbra is 
accomplished by convection.

In the outer sections of penumbral filaments (i.e. further from the umbra) we found maximum blueshifts of up to 3.0\,km\,s$^{-1}$.  
If we assume that of this blueshift is due to the Evershed effect, its projection in the direction of disk 
center and parallel to solar surface gives 8.5\,km\,s$^{-1}$ of radial outflow. This upper limit agrees well 
with results from the simulation by Rempel (2011) which shows Evershed flow speeds above 8\,km\,s$^{-1}$
near $\tau = 1$.

The true strength of the penumbral downflows are expected to be larger than found here for the following reasons.
For the same reasons we expect a number of the locations still seemingly at rest, in reality be 
filled with downflowing gas. 
\begin{enumerate}
\item Incomplete removal of straylight. We have been rather conservative when removing straylight, 
      since the assumed granular contrast of 13\% is smaller than values found by Mathew
      et al. (2009) at 5550\,\AA\, from Hinode and Mart\'{\i}nez Pillet et al. (2011) at 5250\,\AA\, from SUNRISE.
\item A blueshift imposed by the LOS-component of the Evershed flow. The major axis of most of the filaments makes an angle of 
      $\sim 55\,^{\circ}$ with the line	joining the sunspot and the disk center. Thus, we expect radial
      outflow due to Evershed effect to contribute as $V_{LOS} = V_{E}\sin\theta\cos55\,^{\circ} = 
      0.35 V_{E} $ (where $V_{E}$ represents the Evershed velocity) to the LOS velocity. The Evershed 
      flow probably hides a significant amount of the redshift due to downflowing gas 
      (cf. Bharti et al. 2011).
\item The geometry of the filaments. As proposed by Zakharov et al. (2008) the penumbral $\tau=1$
      surface is strongly corrugated, so that we see the disk center side of the filaments more 
      clearly than the other side (see their Fig. 4). Hence, in our sunspot we might expect the 
      downflows to be better visible on the diskward side of a filament where
      they may be partly covered by the filament lying in front of them. This interpretation is 
      supported by the fact that the strongest downflows are found on the diskward side of a filament 
      extending into the umbra, where no other filament can block the view of its edge. 

\end{enumerate}

In summary, we report the discovery of downflows at the edges of  bright penumbral filaments in 
the deep photospheric layers. Such downflows are expected for overturning convection in the penumbra. 

\begin{acknowledgements}
This work has been partially supported by WCU grant No. R31-10016 funded 
by the Korean Ministry of Education, Science and Technology.
The Swedish 1-m Solar Telescope is operated on the island of La Palma
by the Institute for Solar Physics of the Royal Swedish Academy of
Sciences in the Spanish Observatorio del Roque de los Muchachos of the
Instituto de Astrof\'{\i}sica de Canarias. JJ acknowledges a PhD 
fellowship of the International Max Planck Research School on 
Physical Processes in the Solar System and Beyond.
     
\end{acknowledgements}


\end{document}